\documentstyle[twocolumn,pra,aps]{revtex}
\begin{document}
\draft
\author{Li-Bin Fu$^1$\thanks{%
Email: fu{\underline { }}libin@mail.iapcm.ac.cn }, Jie Liu$^2$ and Shi-Gang Chen$^1$}
\address{$^1$Institute of Applied Physics and Computational Mathematics,\\
P.O. Box 8009 (26), 100088 Beijing, China\\
$^2$ Department of Physics, University of Texas, Austin, Texas 78712}
\title{Correlated electron emission in laser-induced nonsequence double ionization
of Helium }
\maketitle

\begin{abstract}
\begin{center}
{\bf Abstract}
\end{center}

In this paper, we have investigated the correlated electron emission of the
nonsequence double ionization (NSDI) in an intense linearly polarized field.
The theoretical model we employed is the semiclassical rescattering model,
the model atom we used is the helium. We find a significant correlation
between magnitude and direction of the momentum of two emission electrons,
and give a good explanation for this striking phenomenon by observing the
classical collisional trajectories. We argue that this correlation
phenomenon is universal in NSDI process, as revealed by the recent
experiment on the argon.
\end{abstract}


The excessive double ionization observed in Helium experiments \cite
{cet3,cet4,cet5} draws much attention to the multiple-electron dynamics in
the laser-atom interaction. In these experiments the single ionization of He
in a linearly polarized field is accurately predicted by the single active
electron (SAE) approximation \cite{cet4}, well described by the
Ammosov-Delone-Krainov (ADK) tunnelling theory \cite{ndi5}. However, the
case of double ionization is more complicated. In the regime of very high
intensities ($I>10^{16}$W/cm$^2$) where strong double ionization occurs, the
double ionization keeps in good agreement with the sequential SAE models as
that in the lower intensities regime($I<10^{14}$W/cm$^2$). The double
ionization deviates seriously from the sequential SAE model and shows a
great enhancement in a ``knee'' regime $[(0.8$---$3.0)\times 10^{15}$W/cm$%
^2] $. This surprising large yields of the double ionization obviously
indicates that the sequential ionization is no longer the dominating process
in this regime and the electron-electron correlation has to be taken into
account. Intense efforts to model the two-electron process of the double
ionization in a laser field have reproduced the main feature of the knee
structure in the double ionization yield as a function of laser peak
intensity and, moreover, yielded quantitative agreement with the experiments
in some cases \cite{n12,fu}.

The physical mechanism behind this nonsequential process is, however, still
debatable. Both the ``shake-off'' model and the ``recollision'' model are
suggested to describe the electron's correlation \cite{cet3,cet5,cfi2,m5}.
However, none of the two nonsequence double ionization (NSDI) mechanisms can
completely explain the experimental observations. For the ``shake-off''
model, it can not give the reason for the decrease in the double ionization
yields as the polarization of the laser field departs from linear \cite
{ee28,ee34,ee35}. In the ``recollision'' model, the returning electrons are
known to have a maximum classical kinetic energy of $\sim 3.2U_p$ $%
(U_p=e^2F^2/4m_e\omega ^2)$, so one can determine a minimum intensity
required for the rescattering electron to have enough energy to excite the
inner electron. But the double ionization yields observed in experiments has
no such an intensity threshold. In fact, the double ionization process is
rather complicated and subtle, both of the two NSDI processes and the
sequential ionization have contributions to it and may dominate in the
different regimes. In another aspect, Becker and Faisal proposed a
``correlated energy sharing'' model to describe the NSDI processes and
nuclei recoil experiment \cite{n12,b2,b3}. The model is based on the
so-called intense-field many-body S-matrix theory derived by a rearrangement
of the usual S-matrix series and include time electron correlation and the
rescattering mechanism.

We have employed the semiclassical model to study the double ionization of
helium in intense linearly polarized field \cite{fu,chengj}. Our
calculations reproduced the excessive double ionization and the
photoelectron spectra observed experimentally both quantitatively and
qualitively, and we argue that the classical collisional trajectories is the
main source of the nonsequence double ionization of helium in the ``knee''
regime.

Recently, the observation of the correlated electron emission in
laser-induced double ionization of argon \cite{nature} provided new insights
into the NSDI process. These authors reported a strong correlation between
the direction and the magnitude of the momenta of two electrons emitted from
an argon atom: the momenta of the two emission electrons tend to have the
same magnitude and sign in the polarization direction. On the theoretical
side, by solving the time-dependent Schr\"odinger equation for two electrons
in three dimensions, Taylor et al \cite{na28} gave that the most of double
ionization probability flux tends to emerge to the same side of the ion.
Similar conclusions have been drawn from one-dimensional model \cite{na2930}.

In this paper, based on the 3D semiclassical rescattering model developed
recently \cite{fu,chengj}, we investigate the dynamical behavior of the
correlated electrons in the double ionization process by analyzing their
classical trajectories. This investigation, as shown later, is very helpful
to understand the physical mechanism behind the momentum correlation. The
model atom we use is the helium, however we argue that our discussions are
available to the other multi-electron atoms, like the argon in recent
experiment\cite{nature}.

Firstly, we briefly present the semiclassical rescattering model adopted in
our calculations. The ionization of the first electron from bound state to
the continues state is treated by the tunnelling ionization theory
generalized by Delone et al. \cite{chj21}. The subsequent evolution of the
ionized electron and the bound electron in the combined Coulomb potential
and the laser fields is described by a classical Newtonian equation. To
emulate the evolution of the electron, a set of trajectories is launched
with initial conditions taken into from the wave function of the tunnelling
electron.

The evolution of the two electrons after the first electron tunnelled are
described by the classical equations (in atomic unit):
\begin{equation}
\frac{d^2{\bf r}_i}{dt^2}=-\nabla (V_n^i+V_{ee})-{\bf F}(t),\;i=1,2.
\label{eq1}
\end{equation}
Here ${\bf F}(t)=F\cos (\omega t)\vec e_z$ is the laser field. The indices $%
i=1$ and $2$ refer to the tunnel ionized and bound electron respectively.
The potentials are
\begin{equation}
V_n^i=-\frac 2{|{\bf r}_i|},\;V_{ee}=\frac 1{|{\bf r}_1-{\bf r}_2|}
\label{pot}
\end{equation}

The initial condition of the tunnelled electron, under the SAE approximation
of He$^{+}$, is determined by a equation including the effective potential
given in Ref. \cite{chj22} and a generalized tunnelling formula developed by
Delone et al. \cite{chj21}. In parabolic coordinates, the Schr\"odinger
equation for a hydrogen-like atom in a uniform field $\epsilon $ is written
(in atomic unit),
\begin{equation}
\frac{d^2\phi }{d\eta ^2}+(\frac{I_{p1}}2+\frac 1{2\eta }+\frac 1{4\eta ^2}+%
\frac 14\epsilon \eta )\phi =0,  \label{sch}
\end{equation}
in which $I_{p1}=-0.9$ $a.u.$ is the negative ionization potential of the
outer electron.

The above equation has the form of the one-dimensional Schr\"odinger
equation with the potential $U(\eta )=-1/4\eta -1/8\eta ^2-\epsilon \eta /8$
and the energy $K=\frac{I_{p1}}4.$

The evolution of the outer electron is traced by launching a set of
trajectories with different initial parameters $t_0$ and $v_{1x0}$, where $%
v_{1x0}$ is the initial velocity perpendicular to the polarization of the
electric field. The initial position of the electron born at time $t_0$ is
given by $x_{10}=y_{10}=0$ , $z_{10}=-\eta _0/2$ from the Eq. (\ref{sch}).
The initial velocity is set to be $v_{1y0}=v_{1z0}=0,$ $v_{1x0}=v_{10}$.
Thus, the weight of each trajectory is evaluated by \cite{chj21}
\begin{equation}
w(t_0,v_{10})=w(0)w(1),  \label{wei1}
\end{equation}
\begin{equation}
w(1)=\frac{\sqrt{2I_{p1}}v_{10}}{\epsilon \pi }\exp (-\sqrt{2I_{p1}}%
v_{10}^2/\epsilon ),  \label{wei2}
\end{equation}
and where $w(0)$ is the tunnelling rate in the quasistatic approximation
\cite{cfi11}.

The initial state of the bounded electron is described by assuming that the
electron is in the ground state of He$^{+}$ with energy $E_2=-2.0\;a.u.$ and
its initial distribution is microcanonical distribution \cite{c28,fu}.

In our calculation, the Eqs. (\ref{eq1}) are solved in a time interval
between $t_0$ and $15T$ by employing the standard Runge-Kuta algorithm$.$
During the first ten optical cycles the electric field amplitude is
constant, and then the field is switched off using a $cos^2$ envelope during
three cycles, and during the last two optical cycles the electrons is free
from the electric field. The wavelength is $\lambda =780$ $nm$, which is so
chosen to match the experiment \cite{cet4,newcc}, and the intensity of laser
is $I=1\times 10^{15}W/cm^2$ which is right at the `knee' region of the
double ionization of helium. $10^6$ or more initial points are randomly
distributed in the parameter plane $-\pi /2<\omega t_0<3\pi /2,$ $v_{1x0}>0$
for the outer electron and in the microcanonical distribution for the inner
electron. The distribution for the ionization electron can be obtained by
making statistics on an ensemble of classical trajectories weighed by the (%
\ref{wei1}). The results have been tested for numerical convergence by
increasing the number of trajectories.

Figure 1 shows the momentum correlation between the two emission electrons
in the double ionization of the present calculations. The horizontal axis
shows the momentum component of the first electron in the direction of
polarization $(P_{1z})$ and the vertical axis the same momentum component of
the second electron $(P_{2z}).$ This figure shows a strong correlation
between the momenta of the two electrons. There is a clear maximum for both
electrons being emitted with the same momentum component in the direction of
polarization axis of about $2.7$ $a.u.$, and emission to opposite half
planes is strongly suppressed, i.e., both two electrons tend to fly to same
side of ion in the direction of polarization. This phenomena has been
observed in the `knee' region for argon \cite{nature}. On the other hand,
from Fig. 1, we see that the maximum momentum of both electron is about $4.5$
$a.u.,$which is consist with the electron-ion coincidence experiment
observation of helium \cite{newcc}, in which the maximum energy of emission
electron in NSDI process is $4U_p,$ since the perpendicular component of
momentum is small, the maximum momentum component in the polarization
direction can be approximate obtained as $P_{z\max }=\sqrt{8U_p}\simeq 4$ $%
a.u..$

An useful alternative perspective on Fig. 2 is obtained by rotating the
distribution by $45^{\circ }$. Then we can get two new distributions. In
Fig. 2(a), we show the distribution of the sum momentum $%
P^{+}=(P_{1z}+P_{2z})$; In Fig. 2(b), we show the distribution of the
difference momentum $P^{-}=$ $(P_{1z}-P_{2z})$. Owing to momentum
conservation, the sum momentum of emission electrons is equal and opposite
to the He$^{2+}$ recoil ion momentum \cite{chen24,chengj}, so its
distribution, as we know, shows a characteristic double-hump structure with
a central minima. A vast amount of literature has devoted to discuss the
double-hump structure of momenta of recoil ions \cite{chen24,chengj}, here
we only pay attention to the correlation between the two emission electrons.

Comparing the Fig.2(a) with Fig.2(b), one finds that the sum momentum at the
peak is about $4.3\;a.u.$, almost above the maximum momentum of each
electron, and the maximum sum momentum is almost $8.7$ $a.u.,$ about twice
of the maximum momentum of one electron; furthermore the distribution width
of difference moment is much smaller than the one of the sum momentum. These
features indicate the momenta parallel to polarization axis of the two
emission electrons likely have the same direction. On the other hand, the
peak of the distribution of difference momentum is at zero, so the momenta
of the two emission electron tend to have the equal magnitude.

To study the origin of the correlation of electrons emission, we show two
trajectories of the electrons in Fig. 3. Fig. 3(a) shows a typical
trajectory of which the difference momentum in polarization direction is
small and fig. 3(b) shows a typical trajectory of which the difference
momentum in polarization direction is nearly the maximum. From these
trajectories, it can be concluded that after the second electron ionized the
momentum transfer received from the laser field is almost identical for both
electron. Because the electron can obtain very high energy from the laser,
the momenta of the two thus be accelerated to the same direction. On the
other hand, this part of the acceleration only adds to the sum of the
momenta of the two electrons, but does not change the difference of the
momenta. Therefore the difference of the momenta is only determined by the
ionization process of the second electron.

As we have discussed before, the second electron is mostly ionized by the
``recollision'' mechanism: the second electron is ionized by a collision
with the tunnelled electron \cite{fu}. Because the collision between the two
electrons is almost instantaneous so that the energy is conserved
approximately when the collision happens. The total energy of the two
electron can be expressed by
\begin{equation}
E=E_T+\left( \frac{P_{1z}+P_{2z}}2\right) ^2+\left( \frac{P_{1z}-P_{2z}}2%
\right) ^2  \label{energy}
\end{equation}
where $E_T$ includes the potentials and the kinetic energy of the
perpendicular parts. To the best of our knowledge, so far the energy
distribution of the returning electron has not been reported, but we believe
that it exhibits a peak at zero energy and decreases as the energy
increasing, on the other hand, only the returning electron which energy is
above the ionization potential of the bound electron $I_{p2}$ can cause the
second electron ionized, so the total energy of the two electrons soon after
the collision occurred is small and its distribution peaks at zero.
Therefore the $P^{-}$ is more likely zero, i.e., the momentum components in
the polarization direction of the two emission electrons are likely equal to
each other. The returning electrons are known to have a maximum classical
kinetic energy of $\sim 3.2U_p,$ therefore the maximum difference momentum
must satisfy $\left( \frac{P_{1z}-P_{2z}}2\right) ^2\simeq 3.2U_p-I_{p2},$
so $|P^{-}|_{\max }=2\sqrt{3.2U_p-I_{p2}}\simeq 4.2\;a.u.$.

On the other hand, since the total energy of the two electrons soon after
the collision occurring is small, the total final energy of the two emission
electrons mostly reflect the energy transfer received from the field. The
field acceleration make the sum momentum increase. The maximum energy for an
electron in the double ionization process is about $4U_p$ \cite{newcc}, from
Eq. (\ref{energy}), assuming the two emission electron have the same energy,
we can obtain the maximum sum momentum $|P^{+}|_{\max }\simeq 8\;a.u.$. This
indicates that the joint acceleration of the electrons in the laser field
clearly dominates over the influence of the electron repulsion, and both
electrons ionized are driven by the laser to the same side.

In conclusion, we have investigated the momentum correlation between
magnitude and direction of the two emission electrons in nonsequence double
ionization. The numerical results on the helium show a significant
correlation on the momentum of the two electrons: the emission electrons
tend to have the same momentum component in the polarization direction. The
phenomena can be directly comprehended from the classical collisional
trajectories. These discussions suggest that the correlated electrons
emission in double ionization process is a semiclassical process. We also
evaluated the width of distribution of the sum and difference momentum.
Because the difference momentum is only determined by the ionization
process, so it is important to verify the dominating process in the ``knee''
regime. Based on the rescattering model, we argue that the maximum
difference momentum of the two emission electrons is $|P^{-}|_{\max }=2\sqrt{%
3.2U_p-I_{p2}}.$ The predictions coincide with the argon experiment \cite
{nature}. We hope our discussions will stimulate the experimental works on
the helium in the direction.

We are indebt to Dr. Jing Chen for reading our paper and proposing many
suggestive opinions. This work was supported by the 973 Project of China.

\section*{Figures caption:}

Fig. 1. Momentum correlation between the two emitted electrons given by
present calculations.

Fig. 2. The distribution of momentum parallel to the polarization axis: (a)
the sum momentum of the two emission electrons. (b) the difference momentum
of the two emission electrons.

Fig.3. Two typical trajectories of the `` recollision'' mechanism: (a) the
case that the difference momentum is small. (b) the case that the difference
momentum is nearly the maximum.


\begin{references}
\bibitem{cet3}  D.N. Fittinghoff, P.R. Bolton, B. Chang, and K.C. Kulander,
Phys. Rev. Lett. {\bf 69}, 2642 (1992).

\bibitem{cet4}  B. Walker, B. Sheehy, L.F. DiMauro, P. Agostini, K.J.
Schafer, and K.C. Kulander, Phys. Rev. Lett. {\bf 73}, 1227 (1994)

\bibitem{cet5}  B. Sheehy, R. Lafon, M. Widmer, B. Walker, L.F. DiMauro,
P.A. Agostini, and K.C. Kulander, Phys. Rev. A {\bf 58}, 3942 (1998)

\bibitem{ndi5}  M.V. Ammosov, N.B. Delone, and V.P. Krainov, Sov. Phys. JETP
{\bf 64}, 1191 (1986)

\bibitem{n12}  A. Becker and F.H.M. Faisal, J. Phys. B {\bf 32}, L335
(1996);Becker and F.H.M. Faisal, J. Phys. B {\bf 29}, L197 (1996)

\bibitem{fu}  Li-Bin Fu, Jie Liu, Jing Chen and Shi-Gang Chen, Phys. Rev. A
{\bf 63}, 043416(2001)

\bibitem{cfi2}  P.B. Corkum, Phys. Rev. Lett. {\bf 71}, 1994 (1993)

\bibitem{m5}  K.C. Kulander, J. Cooper, and K.J. Schafer, Phys. Rev. A {\bf %
51}, 561 (1995)

\bibitem{ee28}  B. Walker, E. Mevel, B. Yang, P. Breger, J.P. Chambaret, A.
Antonetti, L.F. DiMauro, and P.A. Agostini, Phys. Rev. A {\bf 48} R894 (1993)

\bibitem{ee34}  D.N. Fittinghoff, P.R. Bolton, B. Chang, and K.C. Kulander,
Phys. Rev. A {\bf 49}, 2174 (1994)

\bibitem{ee35}  K. Kondo, A. Sagisaka, T. Tamida, Y. Nabekawa, and S.
Watanabe, Phys. Rev. A {\bf 48}, R2531 (1993)

\bibitem{b2}  A. Becker and F.H.M. Faisal, Phys. Rev. A {\bf 59}, R1742
(1999)

\bibitem{b3}  A. Becker and F.H.M. Faisal, Phys. Rev. Lett. {\bf 84}, 3546
(2000)

\bibitem{chengj}  J. Chen, J. Liu, L.B. Fu and W.M. Zheng, Phys. Rev. A {\bf %
63}, 011404(R) (2001)

\bibitem{nature}  Th. Weber, H. Giessen, M. Weckenbrock, G. Urbasch, A.
Staudte, L. Spielberger, O. Jagutzki, V. Mergel, M. Vollmer and R. D\"orner,
Nature 405, 658 (2000)

\bibitem{na28}  K.T. Taylor, J.S. Parker, D. Dundas, E. Smyth, and S.
Vitirito, Laser Phys. {\bf 9}, 98 (1999)

\bibitem{na2930}  M. Lein, E.K.U. Gross and V. Engel, J. Phys. B {\bf 33},
433 (2000); M. D\"orr, Optics Express {\bf 6}, 111 (2000)

\bibitem{chj21}  N. B. Delone, and V. P. Krainov, J Opt. Soc. Am. B {\bf 8},
1207 (1991)

\bibitem{chj22}  L.D. Landau, E. M. Lifishitz, {\it Quantum Mechanics. }%
(Rergamon, Oxford, 1977)

\bibitem{cfi11}  A.M. Perelomov, V.S. Popov and V.M. Teren'ev, Zh. Eksp.
Teor. Fiz. 50, 1393 (1966); M.V. Ammosov, N.B. Delone, and V.P. Krainov,
ibid 91, 2008 (1986)

\bibitem{c28}  J.S. Cohen, Phys. Rev. A {\bf 26}, 3008 (1982)

\bibitem{chen24}  Th. Weber et al., Phys. Rev. Lett. {\bf 84}, 443 (2000)

\bibitem{newcc}  R. Lafon, J.L. Chaloupka, B. Sheehy, P.M. Paul, P.Agostini,
K.C. Kulander, and L.F. DiMauro, Phys. Rev. Lett. {\bf 86}, 2762 (2001)
\end{references}
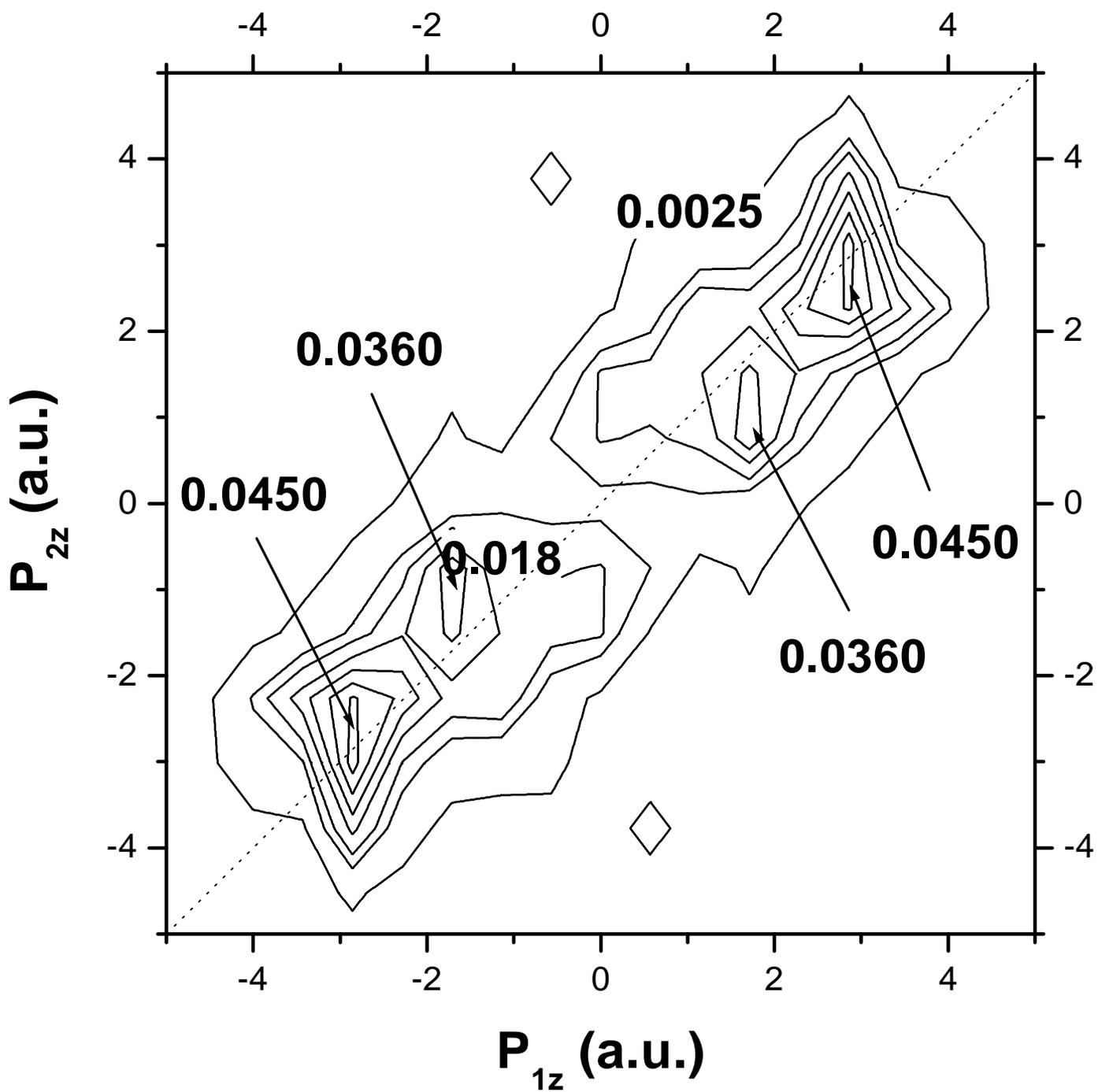
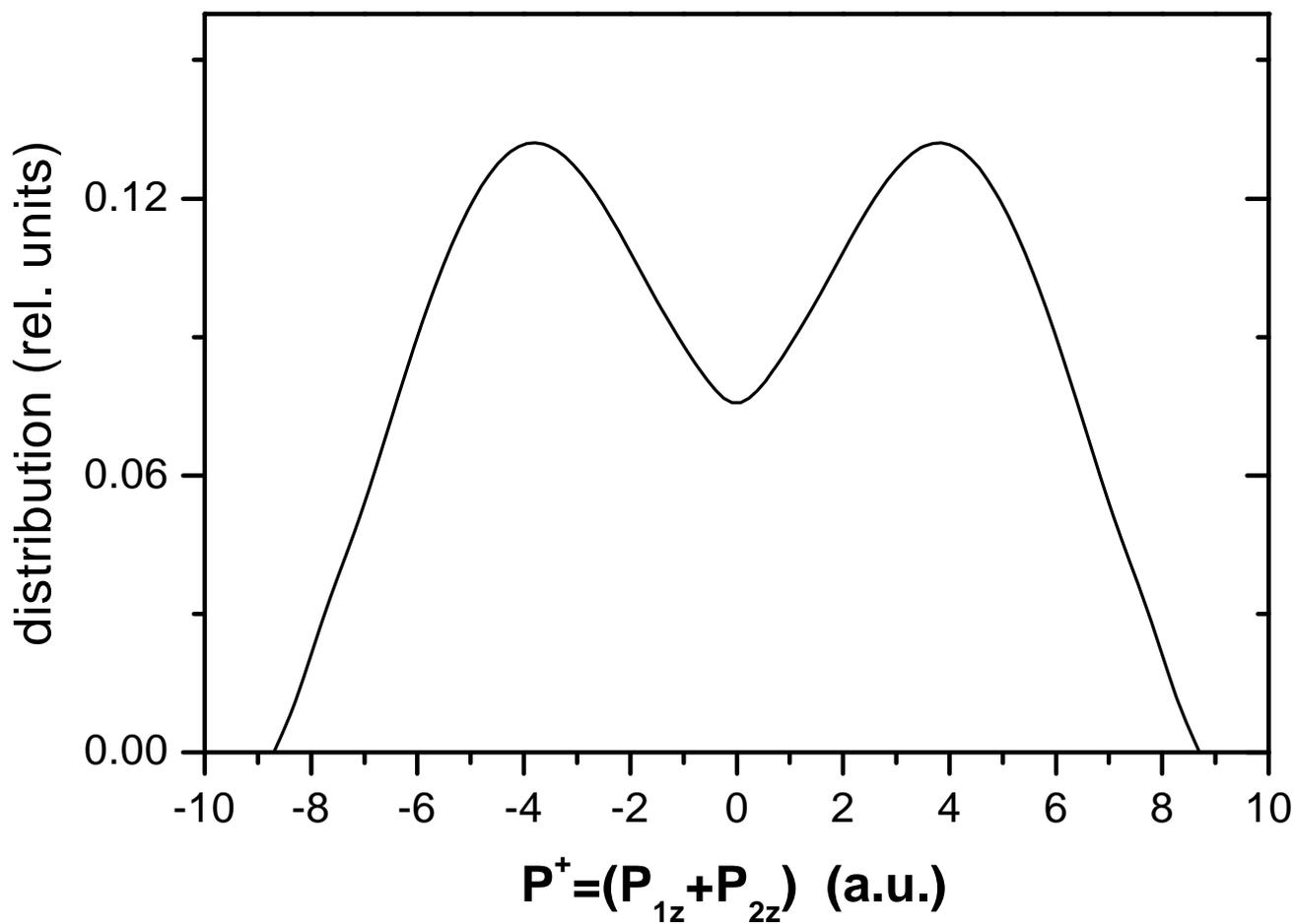
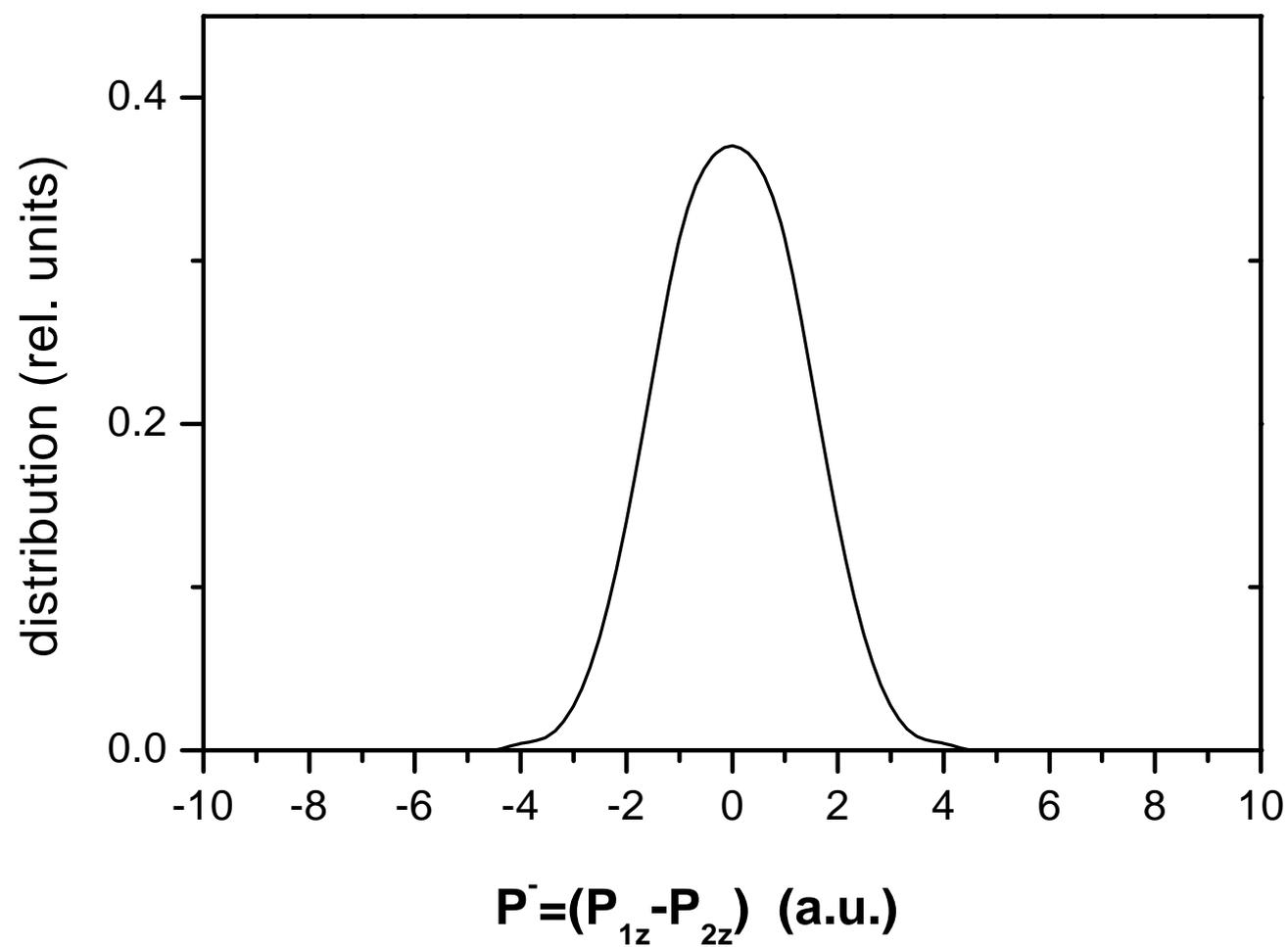
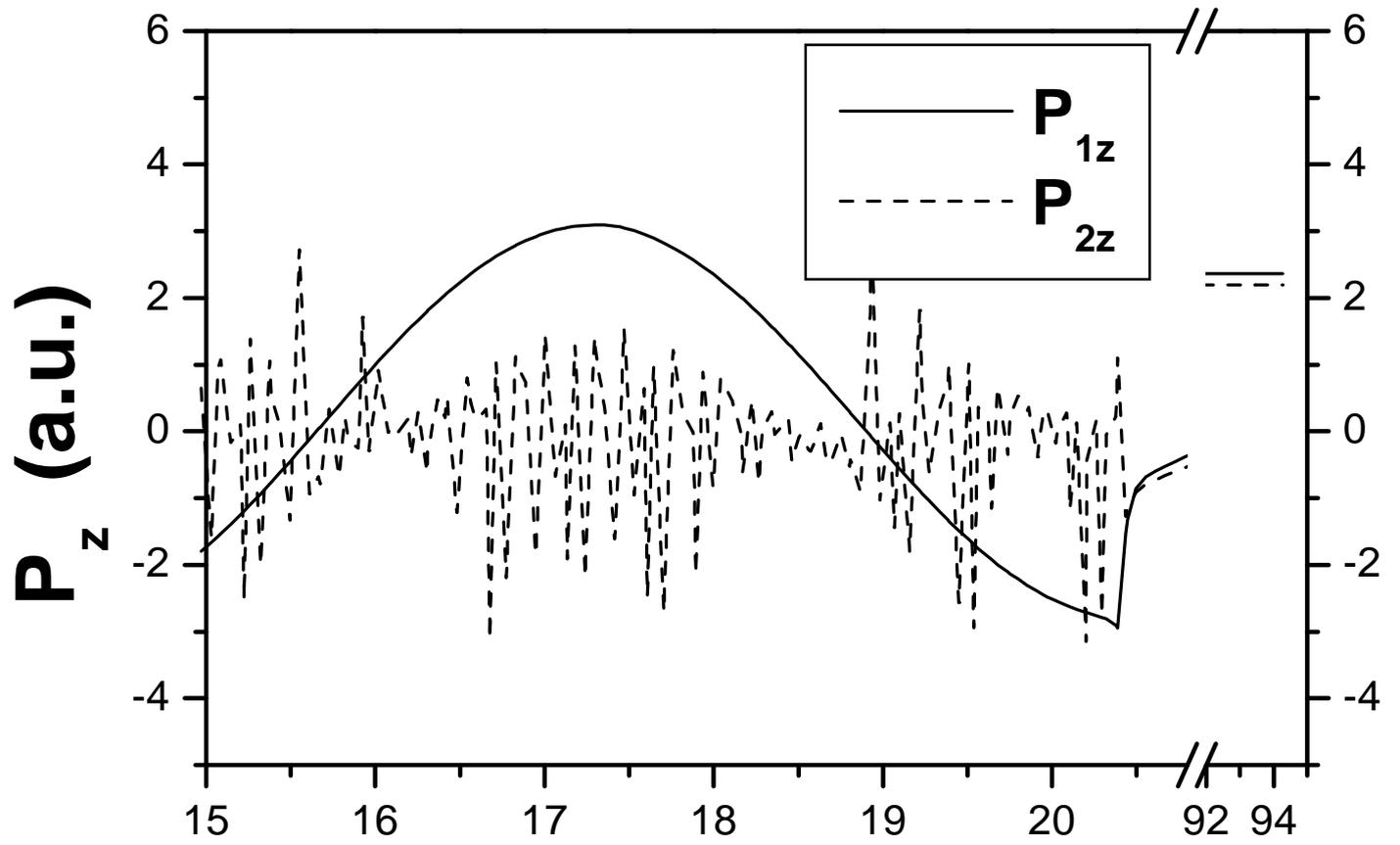
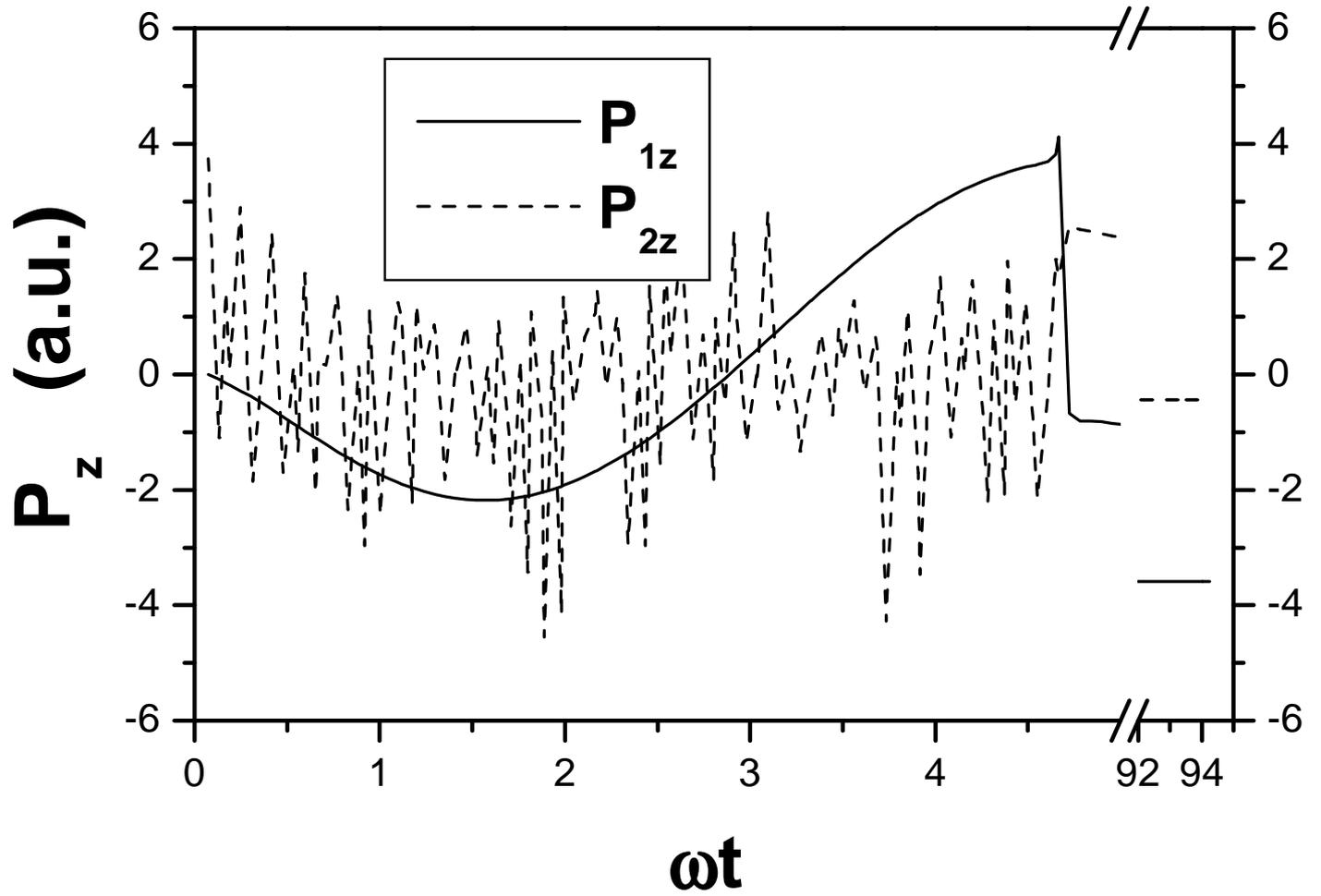
\end{document}